\documentclass[prl,reprint,superscriptaddress,showpacs, hidelinks]{revtex4-1}

\usepackage{graphicx,color}
\usepackage[colorlinks=false]{hyperref}
\usepackage[utf8]{inputenc}

\newcommand{\sz}{\ensuremath{^1\mathrm{S}_0}}
\newcommand{\pz}{\ensuremath{^3\mathrm{P}_0}}

\renewcommand{\vec}[1]{\mathbf{#1}}
\newcommand{\gstate}{\ensuremath{\ket{g}}}
\newcommand{\estate}{\ensuremath{\ket{e}}}
\newcommand{\upstate}{\ensuremath{\ket{\uparrow}}}
\newcommand{\dostate}{\ensuremath{\ket{\downarrow}}}
\newcommand{\eupstate}{\ensuremath{\ket{e\uparrow}}}
\newcommand{\edostate}{\ensuremath{\ket{e\downarrow}}}
\newcommand{\gupstate}{\ensuremath{\ket{g\uparrow}}}
\newcommand{\gdostate}{\ensuremath{\ket{g\downarrow}}}
\newcommand{\ket}[1]{\ensuremath{\left|#1\right\rangle}}
\newcommand{\bra}[1]{\ensuremath{\left\langle#1\right |}}

\begin{document}

\title{Localized magnetic moments with tunable spin exchange in a gas of ultracold fermions}


\author{L.~Riegger}
\author{N.~Darkwah Oppong}
\author{M.~H\"ofer}
\author{D.~R. Fernandes}
\author{I.~Bloch}
\author{S.~F\"olling}
\affiliation{Ludwig-Maximilians-Universit\"at, Schellingstra\ss{}e 4, 80799 M\"unchen, Germany}
\affiliation{Max-Planck-Institut f\"ur Quantenoptik, Hans-Kopfermann-Stra\ss{}e 1, 85748 Garching, Germany}
	
\date{\today}



\begin{abstract}
We report on the experimental realization of a state-dependent lattice for a two-orbital fermionic quantum gas with strong interorbital spin exchange.
In our state-dependent lattice, the ground and metastable excited electronic states of $^{173}$Yb take the roles of itinerant and localized magnetic moments, respectively.
Repulsive on-site interactions in conjunction with the tunnel mobility lead to spin exchange between mobile and localized particles, modeling the coupling term in the well-known Kondo Hamiltonian.
In addition, we find that this exchange process can be tuned resonantly by varying the on-site confinement.
We attribute this to a resonant coupling to center-of-mass excited bound states of one interorbital scattering channel.
\end{abstract}

\maketitle


In many materials, such as transition metal oxides, electrons exhibit an orbital degree of freedom in addition to their spin \cite{Tokura2000}. 
In these systems, electrons in different orbitals give rise to localized and mobile magnetic moments.
Interorbital coupling of theses moments via spin-exchanging interactions leads to the appearance of the Kondo effect, heavy fermion physics, or colossal magnetoresistance \cite{Hewson1993, vandenBrink2004, Coleman2007, Lohneysen2007}.

While ultracold fermionic atoms  have been successfully used to realize a large variety of many-body phenomena \cite{Bloch2012}, two-orbital models such as the Anderson or Kondo models \cite{Hewson1993} have so far not been accessible.
Alkaline-earth-like atoms (AEAs), such as ytterbium and strontium, have been proposed to offer very direct and robust implementations of such systems \cite{Gorshkov2010, Foss-Feig2010, Foss-Feig2010b}, compared to more complex approaches intended for alkali atoms \cite{Paredes2005, Nishida2013, Bauer2013, Bauer2015, Nishida2016}.
The two orbitals can be represented by AEAs in the two lowest-lying states of their electronic spin singlet and triplet manifold, \sz{} (denoted \gstate) and \pz{} (\estate).
As the \estate{} state connects to \gstate{} only via an ultranarrow clock transition, it is stable on typical experimental timescales \cite{Porsev2004}.

Utilizing the different ac polarizabilities of the two states \cite{Dzuba2010} allows for the creation of state-dependent optical lattices (SDLs) \cite{Daley2008,Gerbier2010}. 
In contrast to implementations with alkaline atoms using near-resonant light \cite{Mandel2003,Karski2009,Gadway2010} or field gradient modulation \cite{Jotzu2015}, a SDL for AEAs can be operated at large detuning and low scattering, and preserves the nuclear spin degree of freedom.
The latter arises because the nuclear spin $I$ is strongly decoupled from the electronic states \estate{} and \gstate{} \cite{Scazza2014,Zhang2014}. 
The strong decoupling also preserves the nuclear spin composition during $s$-wave collisions. For isotopes with large $I>1/2$, this can allow for the realization of effective reduced spin models with  SU($N\le2I+1$)-symmetric form \cite{Gorshkov2010, Honerkamp2004, Hermele2009, Corboz2011, Cazalilla2014}.

\begin{figure}[t!]
	\begin{centering}
	\includegraphics[width=\columnwidth]{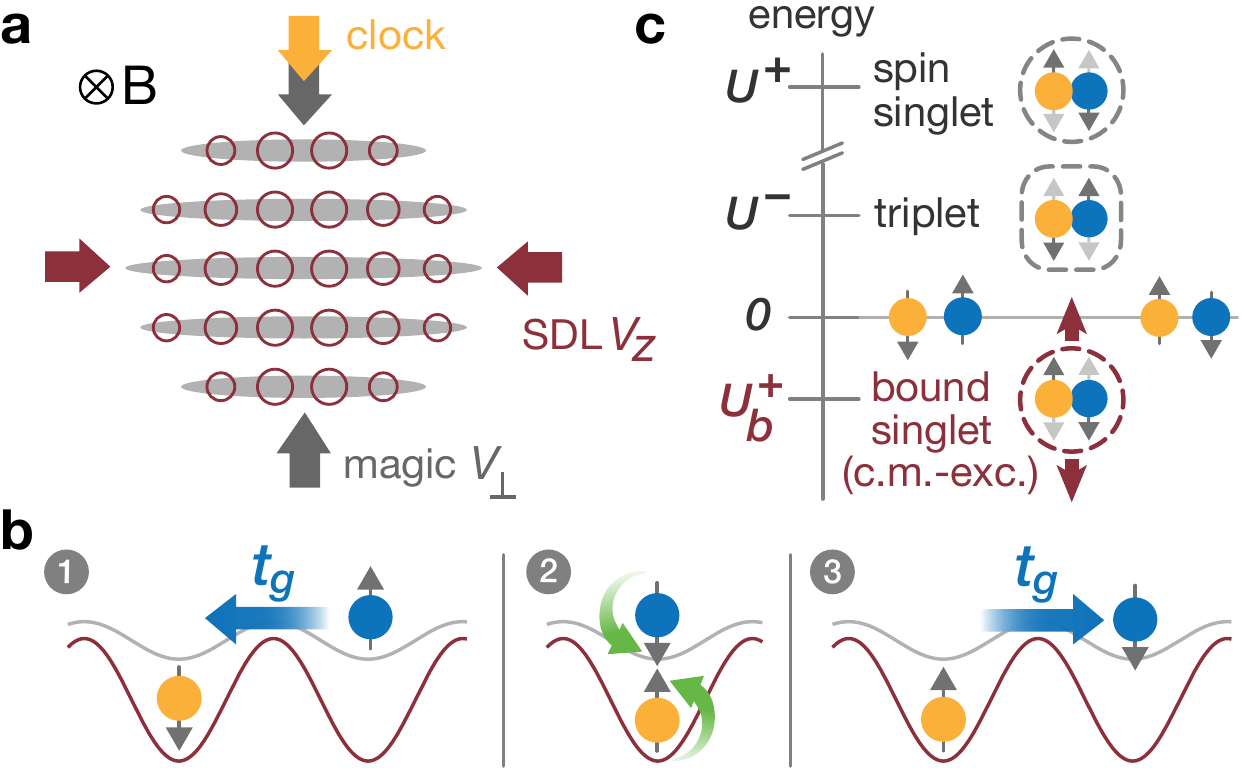}
	\caption{Schematic of the experimental configuration and the spin-exchange mechanism.
	(a) Lattice geometry of isolated quasi-1D tubes with a longitudinal monochromatic SDL. \gstate{} atoms are mobile in a shallow potential (gray) whereas \estate{} atoms are localized in deep lattice wells (red).
	(b) Interorbital spin exchange between \gstate{} atoms (blue) and \estate{} atoms (yellow) on neighboring lattice sites of the SDL consisting of two tunneling processes.
	(c) Energies of the repulsively interacting states \ket{+}, \ket{-} (nuclear spin ``singlet'', ``triplet''), and a c.m.-excited bound state \ket{+_b}, tunable by on-site confinement. All energies are referenced to a pair of noninteracting atoms.}	\label{fig:schematics}
	\end{centering}
\end{figure}

In the presence of a lattice potential, an on-site orbital spin-exchanging interaction emerges in the corresponding single-band Hubbard model \cite{Gorshkov2010}. 
Fermionic $^{173}$Yb ($I=5/2$) stands out for the investigation of orbital magnetism, due to its strong interorbital spin-exchange interaction.
This is a consequence of the unusually large difference in the scattering lengths $a_{eg}^+= 1878(37)\,a_0$ and $a_{eg}^-= 220(2)\,a_0$ \cite{Hofer2015} of atom pairs in symmetric and antisymmetric superposition states of $\ket{g}$ and $\ket{e}$.   
The resulting fundamental exchange coupling is ferromagnetic and large compared to the Fermi energy, temperature and intraorbital interactions, both in deep state-independent potential wells and in free space \cite{Scazza2014, Cappellini2014}. 

Spin-exchanging interactions without a lattice or without a mobile orbital have previously been implemented in ultracold atom experiments \cite{Anderlini2007, Scazza2014, Cappellini2014}.
In the present work, these interactions are combined with orbital state-dependent mobility. 
As a result, a second-order spin-exchange mechanism arises that couples  mobile to localized magnetic moments analogous to the Kondo coupling arising from the single impurity Anderson model. 
Furthermore, we show that the exchange can be tuned resonantly via the external confinement.


In our experiment, the atoms are confined in an array of independent quasi-1D traps, created by two perpendicular state-independent lattices operated at the magic wavelength $\lambda_{\mathrm{m}}=759.4$\,nm \cite{Barber2008} [see Fig.~\ref{fig:schematics}(a)]. 
Along the remaining, longitudinal direction, we superimpose a state-dependent optical lattice. 
At our chosen wavelength of $\lambda_\mathrm{SDL}=670.0$\,nm, the ac polarizability of \estate{} atoms is $3.3$ times larger than for the \gstate{} atoms \cite{SM}, independent of the nuclear spin component $m_F$.
We operate the SDL at intermediate depths, where both states are in the tight-binding regime. 
However, the \estate{} atoms are effectively localized on experimentally relevant timescales  while the \gstate{} atoms retain their mobility.

In analogy to the well-known Kondo model, we consider a spin-exchanging collision, where a mobile \gstate{} particle is scattered off a localized \estate{} state moment. 
In the tight-binding picture of the state-dependent Hubbard model, for two particles initially on separate sites, this requires tunneling of the \gstate{} particle to the \estate{} site and back, as depicted in Fig. \ref{fig:schematics}(b).
In this super-exchange-like process, the energy of the interacting intermediate state depends on the spin configuration of the two particles.
We employ two nuclear spin components \upstate{} or \dostate{} ($m_F = \pm 5/2$) to realize the spin degree of freedom. This can be considered an effective spin-1/2 system due to the absence of spin-changing processes on relevant timescales.
Neglecting bound states of the atom pair, the lowest-energy on-site pair state is $\ket -=\frac{1}{2}(\ket{eg}-\ket{ge})(\ket{\uparrow\downarrow}+\ket{\downarrow\uparrow})$ with interaction energy $U^-$, denoted as ``spin triplet.'' The interaction energy $U^+$ of the state $\ket +=\frac{1}{2}(\ket{eg}+\ket{ge})(\ket{\uparrow\downarrow}-\ket{\downarrow\uparrow})$, denoted by ``spin singlet,'' is much larger [see Fig.~\ref{fig:schematics}(c)].
With both energies large compared to the hopping rate $t_g$, the dominant spin-triplet state leads to an effective ferromagnetic spin-exchange coupling $J_\mathrm{ex} \approx - t_g^2/{U^-} < 0$.

\begin{figure}[t!]
	\begin{centering}
	\includegraphics[width=\columnwidth]{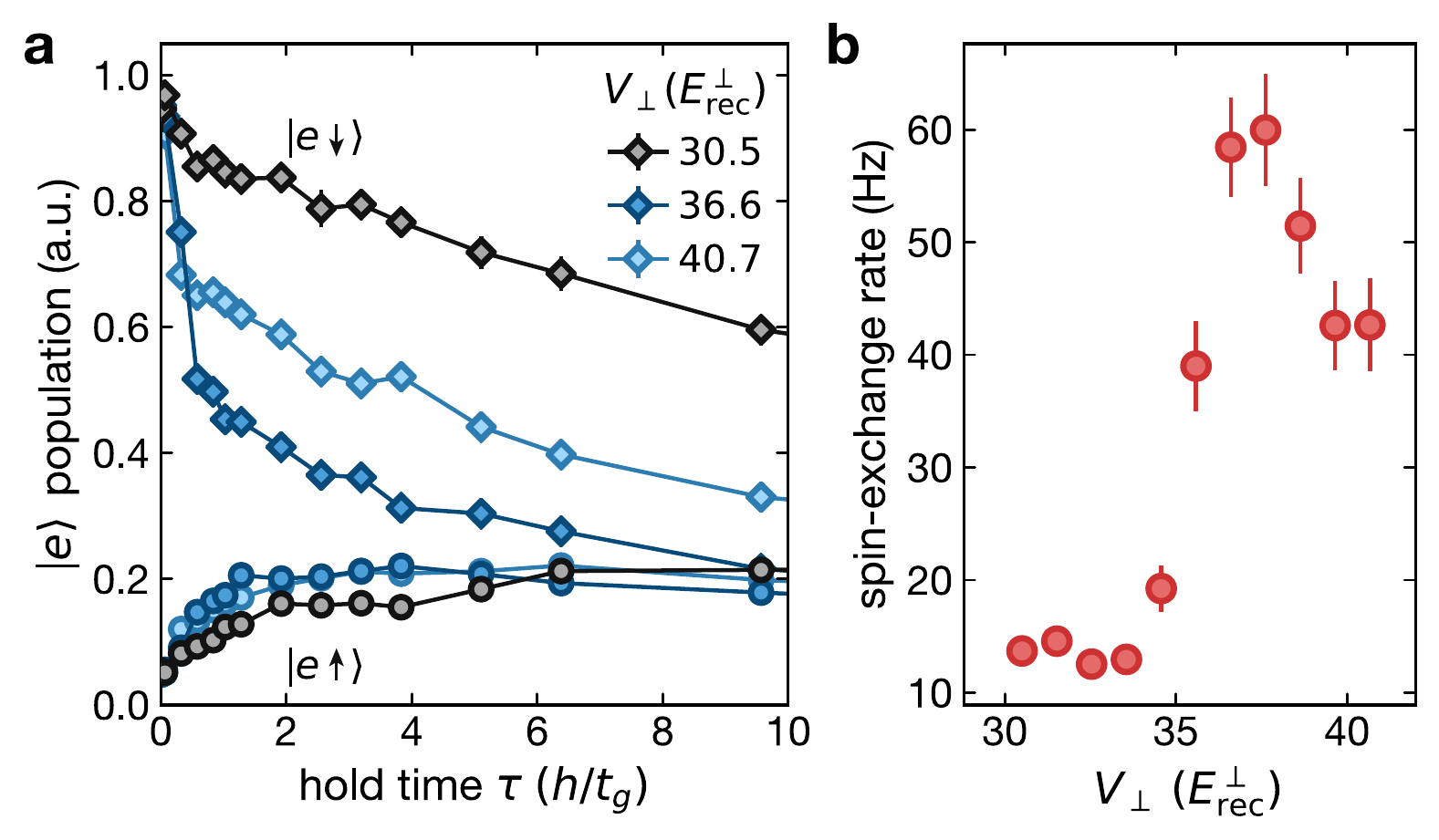}
	\caption{Spin-exchange dynamics and dependence on perpendicular confinement.
	 (a) The time dependence of the \edostate{} (\eupstate{}) population relative to the initial, total \estate{} atom number is shown as diamonds (circles). 
	 Each point is the average of at least four individual measurements.
	 Data points less than $0.2\,h/t_g$ apart are binned to reduce visual clutter.
	 Error bars indicate the standard error of the mean.
	 (b) Rate of increase of the \eupstate{} fraction in the \estate{} population, determined by a linear fit to the short-time dynamics up to one tunneling time.
	 Error bars indicate the $1\sigma$ uncertainties of the fits.
	All atom numbers have been extracted from an image area covering 33\% of  the atomic cloud size in the center of the trap.
	For all measurements, $V_z = 5.7 \,E_\mathrm{rec}^z$.
	}
	\label{fig:spinexchange-dynamics}
	\end{centering}
\end{figure}

To probe this spin-exchange process experimentally, we investigate the spin-equilibration dynamics starting from a mixture of \edostate{} and \gupstate{} atoms on initially singly occupied lattice sites. 
The mobility of the ground state atoms then enables them to reach \edostate{} sites, leading to spin-flip processes $\edostate\gupstate \rightarrow \eupstate\gdostate$ described above.
For the initial state preparation, a balanced two-spin ($m_F=\pm 5/2$) degenerate \gstate{} Fermi gas with  $T\approx0.2\,T_{\rm F}$ is loaded into a deep 3D lattice ($V_z = 8 \,E_\mathrm{rec}^z$ and $V_\perp = 45...50 \,E_\mathrm{rec}^\perp$) at a low atom density, such that double occupancy of sites is substantially suppressed.
Here, the longitudinal (perpendicular) lattice depth $V_z$ $(V_\perp)$ is given in units of the longitudinal (perpendicular) lattice recoil energy $E_\mathrm{rec}^z/h = 2.57\, \mathrm{kHz}$ ($E_\mathrm{rec}^\perp/h = 2.00\, \mathrm{kHz}$) for \gstate{} atoms.
Applying a magnetic bias field of $20\,$G allows us to then selectively transfer the atoms in state \gdostate{} to \edostate{} with a fast $\pi$ pulse on the clock transition, finishing the initial state preparation.

In order to initiate the spin-exchange dynamics, the perpendicular magic-wavelength lattice and the longitudinal SDL depth are rapidly lowered to the values of interest, $V_{\perp}$ and $V_{z}$. 
Simultaneously, the magnetic field is ramped to a small bias value of $1$\,G.
During a variable hold time $\tau$, the \gstate{} atoms are then free to move longitudinally and interact with stationary \estate{} atoms. 
Finally, we image both excited-state spin components separately to measure the evolution of the spin populations \cite{SM}. 

Figure~\ref{fig:spinexchange-dynamics}(a) shows the resulting time dependence of the spin-exchange dynamics in a SDL of intermediate depth ($V_z = 5.7 \,E_\mathrm{rec}^z$ with tunneling rate $t_g/h = 140\,\mathrm{Hz}$ for \gstate{} and $t_e/h = 8\,\mathrm{Hz}$ for \estate).
In the course of several tunneling times for the \gstate{} atoms, the \eupstate{} component is populated through spin exchange and the spin distribution in both orbitals equilibrates. 
In Fig.~\ref{fig:spinexchange-dynamics}(a), the black curve is attributed to the triplet-state dominated spin-exchange model described above, and in this particular case we observe a redistribution on a timescale  $\sim h/J_\mathrm{ex} = 26\,h/t_g$.

In contrast to the behavior expected from this model, we find that the spin-exchange rate can be strongly modified when choosing certain values of the transverse confinement [blue curves in Fig.~\ref{fig:spinexchange-dynamics}(a)].
This is illustrated in Fig.~\ref{fig:spinexchange-dynamics}(b), where a rate is determined with a linear fit to the initial dynamics of the \eupstate{} fraction.
This exchange rate is resonantly enhanced with a maximum around $V_{\perp} = 37\, E_\mathrm{rec}^\perp$ and displays an asymmetric resonance profile. 
Figure~\ref{fig:spinexchange-dynamics}(a) shows that, together with the enhanced spin-exchange rate, also the loss rate of both excited-state species is amplified (see the discussion later in the Letter). 

\begin{figure}[t!]
	\begin{centering}
	\includegraphics[width=1\columnwidth]{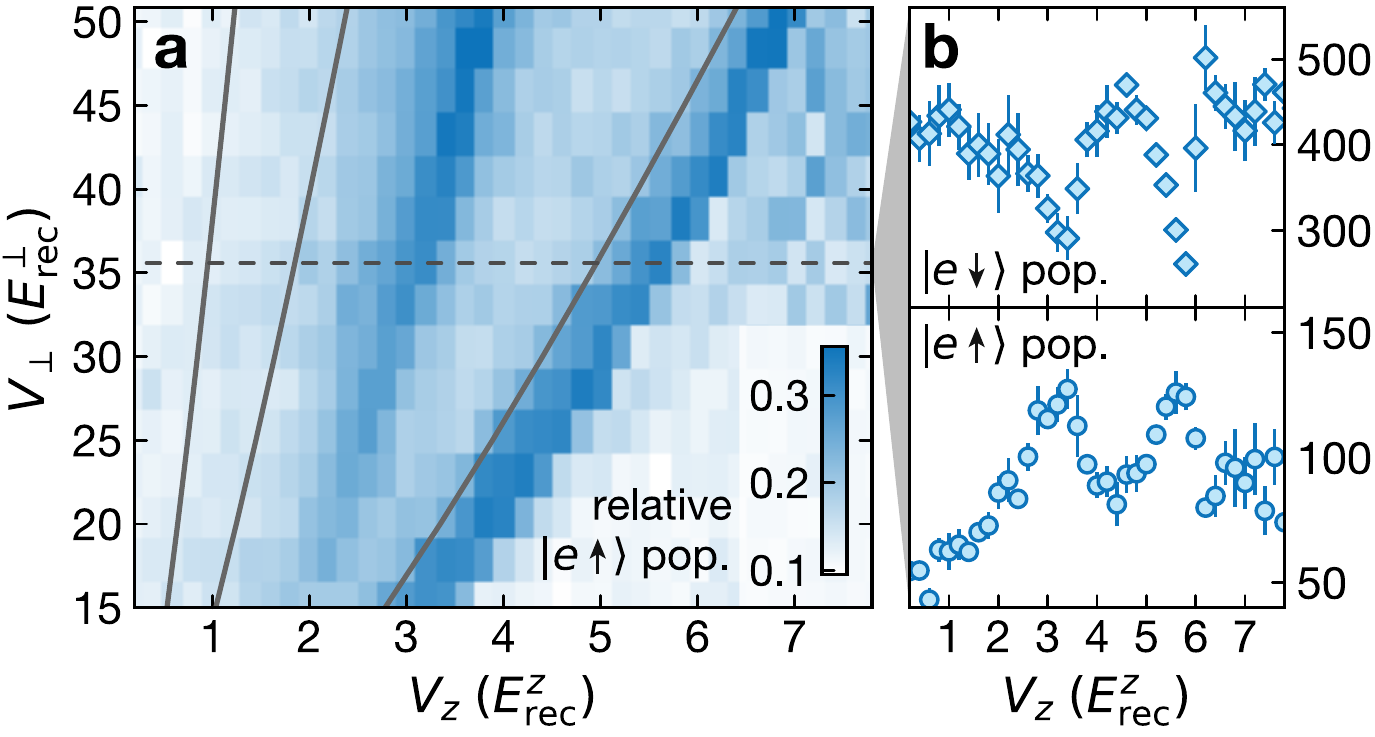}
	\caption{(a) Normalized \eupstate{} atom number after hold time $\tau \propto  t_g^2/U$.
	The hold time is chosen such that $\tau = 2\,t_g$ for $V_z = 6 \,E_\mathrm{rec}^z$ and $V_\perp = 30\,E_\mathrm{rec}^\perp$.
	Two resonance branches appear in the spin-exchange rate as a function of the SDL depth and perpendicular confinement.
	Solid gray lines indicate the predicted resonance position from the anisotropic harmonic oscillator model (second, third and fourth longitudinal c.m. excitation from right to left).
	(b) Cut through the data in (a) at $V_\perp = 35.6\,E_\mathrm{rec}^\perp$ (dashed line) show non-normalized \edostate{} (diamonds) and \eupstate{} (circles) atom counts. 
	All atom numbers are extracted from an image area covering 33\% of the atomic cloud size in the center of the trap. 
	}
	\label{fig:spinexchange-resonances}
	\end{centering}
\end{figure}

To probe the structure of this spin-exchange resonance, we map out the resonance position dependence on both perpendicular as well as longitudinal confinement.
In Fig.~\ref{fig:spinexchange-resonances}(a), we show the growing \eupstate{} fraction after a short hold time $\tau \propto  t_g^2/U$. 
This scaling of $\tau$ is motivated by the super-exchange model for the spin-exchange process.
Varying both confinement strengths, we identify two resonance branches in the experimentally accessible parameter regime. 
A cut along $V_\perp = 35.6\,E_\mathrm{rec}^\perp$ [see Fig.~\ref{fig:spinexchange-resonances}(b)] shows similar properties of the two observed branches.

\begin{figure}[t!]
	\begin{centering}
	\includegraphics[width=1\columnwidth]{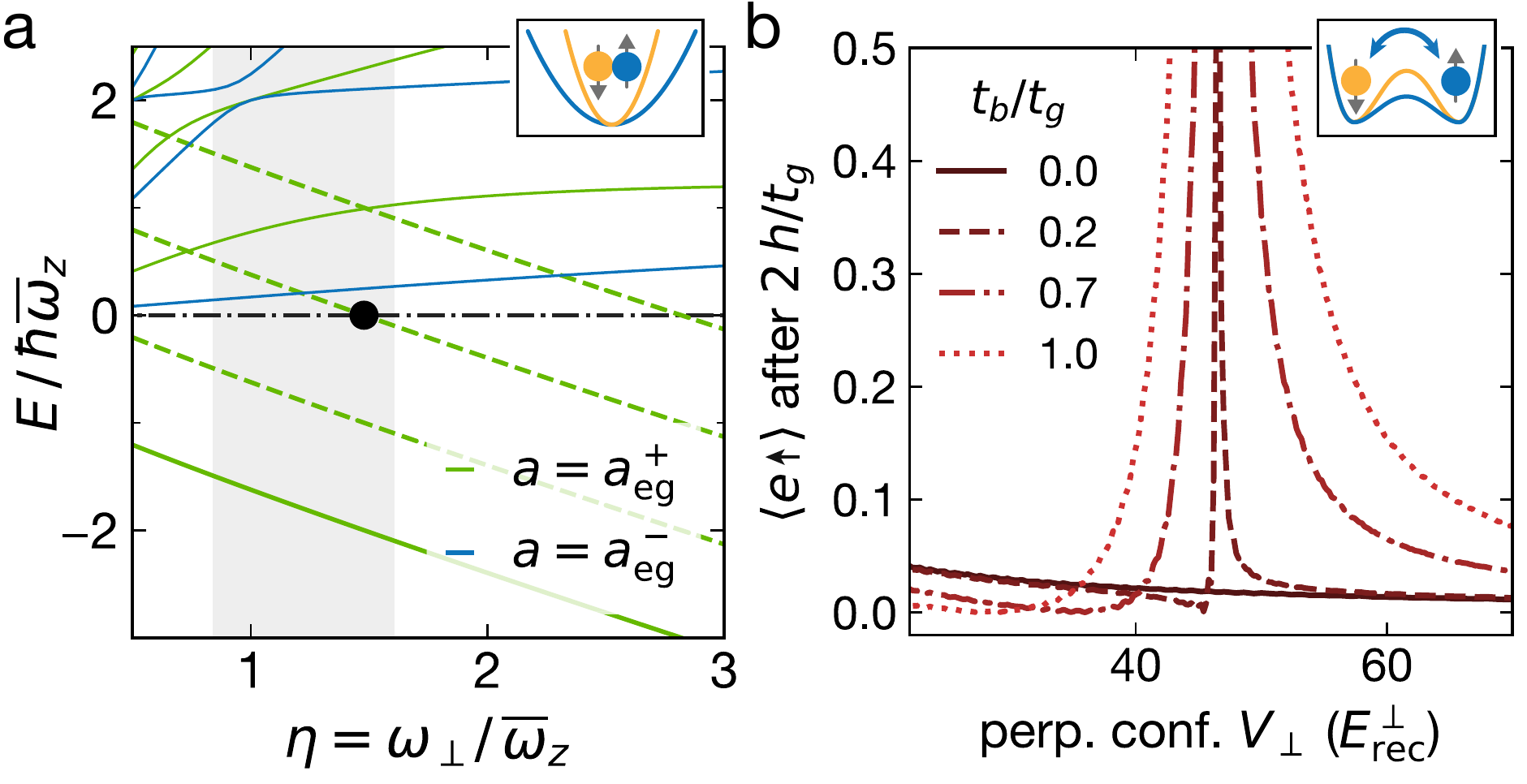}
    \caption{
    (a) On-site eigenenergies of the spin singlet (green) and triplet (blue) state, obtained from the harmonic oscillator model for varying anisotropy $\eta$.
    Solid lines denote states without c.m. excitation.
    The c.m. excited singlet bound states (dashed lines) come into resonance with the zero-point energy of noninteracting particles (dash-dotted line).
    One resonance (black circle) occurs for fixed $V_z=6\, E_\mathrm{rec}^z$ in our available range of $V_\perp$ (gray area).
    (b) Spin-exchange resonance shape from the state-dependent double-well model:
    expectation value of the \eupstate{} fraction after evolution from the  initial state $\ket{e\downarrow}_L\ket{g\uparrow}_R$ for a time $2 h/t_g$ and varying $V_\perp$ around the resonance indicated in (a).
    The resonance width depends on the coupling strength $t_b$ into $\ket{+_b}$. 
    Coherent oscillations from resonant coupling close to the resonance center are not shown for visual clarity. 
	}
   	\label{fig:comrel-resonances}
   	\end{centering}
\end{figure}

A spin-exchange coupling that depends on the perpendicular confinement due to confinement-induced resonances in a quasi-1D system has been proposed in Ref.~\cite{Zhang2016}.
For the achievable lateral confinements in our system, we can rule out the observation of a 1D confinement-induced resonance. 
However, both available scattering channels are repulsive and therefore feature bound states below the energy of non-interacting particles [see Fig.~\ref{fig:comrel-resonances} (a)]. 
Together with the strong longitudinal confinement dependence of the exchange, this suggests a confinement-induced resonance effect in the combined transverse and longitudinal potential. 
One theory for this type of effect has been described in Ref.~\cite{Cheng2017}, using a setting in which a harmonically localized impurity interacts with atoms in a quasi-1D potential.

Here, we concentrate on the tight-binding limit where both orbital states exhibit strong (but different) on-site confinements.
$^{173}\mathrm{Yb}$ features a very shallow bound state of the $a_{eg}^+$ scattering channel with an energy in the range of a few longitudinal center-of-mass (c.m.) excitation energies. 
Therefore, the lowest c.m. excited bound states can come into resonance with two noninteracting particles [see Fig.~\ref{fig:schematics}(c)].
When this happens at certain combinations of transverse and longitudinal confinement, this ``singlet'' bound state $\ket{+_b}$ of the effective spin-1/2 system becomes available as an additional intermediate state for the spin-exchange process. 
As the energy of this intermediate state depends on the tunable optical potentials, the overall process becomes controllable in strength and should even allow for an antiferromagnetic exchange coupling.


The qualitative behavior of the spin-exchange resonances can be explained by means of a simplified two-particle model for the strong interorbital interactions in the presence of a state-dependent confinement.
In the limit of deep lattices, we treat the on-site physics using harmonic oscillator potentials. 
While the radial trapping frequency $\omega_{\perp}$ is the same for the two particles, 
the longitudinal frequencies $\omega_{z,g}$ and $\omega_{z,e}$ along the SDL depend on the orbital $e/g$. 

This setting of anisotropic and mixed confinement for an $eg$ pair can be approximately mapped to a sum of effective harmonic oscillator potentials in reduced c.m. and relative (rel.) coordinates, $\vec R = (\rho_\mathrm{CM}, Z)$ and $\vec r = (\rho,z)$:
\begin{eqnarray}
\hat{H}_\mathrm{c.m.} & = & -\vec{\nabla_{R}}^{2}/2+\left[\eta^{2}\rho_\mathrm{CM}^{2}+Z^{2}\right]/2\\
\hat{H}_\mathrm{rel.} & = & -\vec{\nabla_{r}}^{2}/2+\left[\eta^{2}\rho^{2}+z^{2}\right]/2+V_\mathrm{int}(\mathbf{r}).
\end{eqnarray}
Moreover, there is a coupling term $\hat H_\mathrm{mix} = (\Delta\omega_z^{2}/\overline{\omega}_{z}^2) Zz /2 $ between the rel. and c.m. coordinate with $\Delta \omega_z^2 = \omega_{z,e}^{2}-\omega_{z,g}^{2}$ caused by the mixed confinement \cite{Deuretzbacher2008}.
Our polarizability ratio of $p=3.3$ for the SDL causes a considerable coupling of $\Delta \omega_z^2 / \overline{\omega}_z^2 = 2(p-1)/(p+1) \approx 1$.
Here, all energies are given in units of an effective longitudinal trapping frequency $\hbar \overline{\omega}_z$ with $\overline{\omega}_z^2 =(\omega_{z,g}^{2}+\omega_{z,e}^{2})/2$, and all lengths are in units of the corresponding harmonic oscillator lengths. 
The ratio $\eta = \omega_\perp/\overline{\omega}_z$ characterizes the anisotropy of an individual lattice site and is in the range $0.7<\eta<2.3$.
Interactions are modeled by a regularized contact potential $V_\mathrm{int} = 4\pi a_{eg}^\pm \delta(\vec r) \frac{\partial }{\partial r}{r}$ in the rel. coordinate where the scattering length depends on the state $\ket{+}$ or $\ket{-}$ of the $eg$ pair.
For more details on the model parameters, see Ref.~\cite{SM}.

In a first approximation, we set $\hat H_\mathrm{mix}=0$ and neglect the trap anharmonicity, such that the rel. and c.m. coordinate of the harmonic oscillator decouple. 
The problem can then be solved exactly for arbitrary $\eta$ \cite{Idziaszek2005}.
Figure~\ref{fig:comrel-resonances} depicts the relevant parts of the resulting spectrum, the interacting states in the rel. coordinate Hamiltonian (solid lines) as well as the lowest c.m. excitations (dashed lines) of $\ket{+_b}$.
All energies are shown relative to the zero-point energy $E_0 = (2\eta + 1)\hbar\overline{{\omega}}_{z}$ of the noninteracting harmonic oscillator which we use as the entrance energy of the scattering process starting with two particles on separate sites. 
The spin-singlet bound state $\ket{+_b}$ is loosely bound and the lowest c.m. excitations can be resonant with $E_0$, as shown in Fig.~\ref{fig:comrel-resonances} (a).
For a fixed longitudinal confinement $V_z$, several resonances occur as a function of $\eta$, varied by modifying the transverse confinement $\omega_{\perp}$. 
Overlaying the zero-crossing positions (gray lines) with our experimental data in Fig.~\ref{fig:spinexchange-resonances}, we find that the resonances caused by the longitudinal c.m. excitations roughly match the functional dependence of the observed spin-exchange resonances. 
The validity of the harmonic oscillator model is improved with the SDL depth and for lower c.m. excitations. 
Indeed, for the branch with two c.m. excitations, also the position of the resonance is approximately reproduced.

An effect of the resonant singlet channel on the spin-exchange coupling requires a finite coupling $t_b$ for the process of a \gstate{} atom tunneling to a neighboring site and forming a c.m.-excited bound state with an \estate{} atom. 
This coupling should be enhanced through the on-site c.m.-rel. motion coupling.
 
The total spin-exchange coupling $J_\mathrm{ex}$ for a given confinement results from the combined effect of all available exchange channels.
In particular, for configurations in which the bound singlet channel provides the largest contribution to the exchange, the resulting effective $J_\mathrm{ex}$ can even switch sign between ferro- and antiferromagnetic exchange coupling.

To investigate the influence of the off-site bound state coupling, we consider a double-well model with two particles that incorporates the on-site interactions discussed above, as well as the mobility of the \gstate{} atoms \cite{SM}. 
Figure~\ref{fig:comrel-resonances}(b) shows the expectation value of finding an \eupstate{} atom after letting an initial state $\ket{e\downarrow}_L\ket{g\uparrow}_R$ on the two sites ($L,R$) of the double well  coherently evolve for a time $2\,h/t_g$. 
If the c.m.-excited bound state $\ket{+_b}$ is neglected ($t_b = 0$, solid line), tuning of the perpendicular confinement induces only weak variations in the dynamics, which are then dominated by the spin triplet $\ket{-}$ intermediate state. 
In contrast, for finite couplings $t_b>0$, a resonance in the spin exchange occurs, caused by the intermediate state $\ket{+_b}$. 
The width of the resonance becomes comparable to the one observed experimentally in Fig.~\ref{fig:spinexchange-dynamics} for $t_b$ on the order of $t_g$.
Also, the asymmetry with a steep flank on the side of low $V_\perp$ is reproduced. 
In the model, it is caused by the competition of ferromagnetic and antiferromagnetic coupling, leading to a zero crossing of the effective spin-exchange coupling, with an antiferromagnetic sign of the effective coupling term between the resonance and the minimum position in Fig.~\ref{fig:comrel-resonances}(b).

Despite the qualitative agreement, the two-particle model is not suited to fully describe the dynamics in our experiment for configurations close to the scattering resonances.
First, the description does not include the loss of atoms, as observed in Fig.~\ref{fig:spinexchange-dynamics}(a).
The resonant structure of the loss feature implies that it is related to the coupling to c.m.-excited $eg$ pairs.
For example, collisions with additional mobile particles could lead to the relaxation of these shallow bound states into more deeply bound molecular states, or to the deexcitation of the c.m. motion, inducing particle loss from the trap. 
However, we observe that relative to the \gstate{}-population loss, \estate{} loss is significantly more pronounced than expected for a simple $eg$-pair process \cite{SM}. 
Clearly, the loss feature is detrimental to the stability very close to the resonances in the current configuration.
These areas of the parameter space will need to be avoided for many-body physics implementations without further modifications. 
A more detailed model including such processes could help to minimize the effect.

Second, the nature of the coupling between c.m. and rel. motion is treated only on a simplified level.  
In particular, a more complete model of the coupling should lead to avoided crossings between the eigenenergy branches, affecting the resulting resonance positions. 
A precise numerical treatment of the optical lattice Hamiltonian, similar to Ref.~\cite{Buechler2010}, would be required to determine more exact values for the effective Hubbard parameters.


In conclusion, we have implemented a two-orbital lattice system with both mobile and localized orbital.
Interorbital spin exchange between the mobile and localized moments is observed in this configuration.
The orbital lattice implementation is general for all AEA systems for a suitable choice of lattice wavelength. With appropriate filling factors of localized spins, it can be used for the realization of both Kondo- and Kondo lattice-type models \cite{Gorshkov2010}.
The spin-exchange coupling can be widely tuned in the vicinity of bound state resonances, in particular due to the large scattering length $a_\mathrm{eg}^+$ of $^{173}$Yb.
Our novel tuning mechanism as well as previously proposed schemes \cite{Gorshkov2010, Nakagawa2015} rely on optical potentials for rapidly modifying the exchange coupling. 
This is advantageous for the investigation of nonequilibrium spin dynamics \cite{Hackl2009, Nuss2015}. 
The SDL and the tuning implementation work independently of the nuclear spin, and both are expected to preserve the SU($N$) symmetry of the interactions. 
Using more than two nuclear spin components of $^{173}\mathrm{Yb}$ in the SDL could directly enable the realization of the Coqblin-Schrieffer model \cite{Coqblin1969, Hewson1993, Gorshkov2010, Kuzmenko2016}.


\begin{acknowledgements}
We acknowledge the valuable and helpful discussions with Hui Zhai, Eugene Demler, Marton Kanász-Nagy, Peng Zhang, Meera M. Parish, Jesper Levinsen, Jan von Delft, Seung-Sup Lee and Frank Deuretzbacher.
This work was supported by the European Research Council through the synergy grant UQUAM and by the European Union's Horizon 2020 funding. 
N.D.O. acknowledges funding from the International Max Planck Research School for Quantum Science and Technology.
\end{acknowledgements}

\renewcommand{\thefigure}{S\arabic{figure}}
\setcounter{figure}{0}
\renewcommand{\theequation}{S.\arabic{equation}}
\setcounter{equation}{0}
\renewcommand{\thesection}{S.\Roman{section}}
\setcounter{section}{0}
\renewcommand{\thetable}{S\arabic{table}}
\setcounter{table}{0}

\onecolumngrid
\newpage

\begin{center}
\noindent\textbf{\large{Supplemental Material}}\
\bigskip
\end{center}

\section{Experiment sequence details}

The ratio $p$ of the AC-polarizabilities of \estate{}- and \gstate{}-atoms in the state-dependent lattice (SDL), and thereby of the lattice potential depths, has been determined by means of independent parametric heating of both species in deep lattice systems. 
At a wavelength of 670.0\,nm, we obtain $p=3.3(1)$, in agreement with the theoretical prediction in \cite{Dzuba2010Supp}.
For the lattice depth range of $V_z = (3 ... 8)E_\mathrm{rec}^z$ used in the experiment, this results in a bandwidth range of (1.2 ... 0.3)\,$h\, \mathrm{kHz}$ and (0.22 ... 0.01)\,$h\, \mathrm{kHz}$ for \gstate{} and \estate{}, respectively.

One of the two perpendicular, magic-wavelength lattice beams ensures Lamb-Dicke conditions for the coaligned 578.4\,nm excitation beam for the clock transition.

We expect and have verified experimentally that, in the limit of strong perpendicular confinement, the position of the spin-exchange resonances only depends on the geometric mean of the individual perpendicular lattice depths $V_\mathrm{x}$ and $V_\mathrm{y}$.
Thus, the perpendicular lattice depth provided in the main text is obtained through $V_\perp = \sqrt{V_\mathrm{x}V_\mathrm{y}}$.
The measurements have been performed with a tiny anisotropy of 1.7\% in the perpendicular trapping frequencies.
This is due to a small mismatch in the experimental calibration of the perpendicular lattice depths $V_\mathrm{x}$ and $V_\mathrm{y}$, determined in an independent measurement.

\begin{figure}[htb]
	\begin{centering}
	\includegraphics[width=0.6\columnwidth]{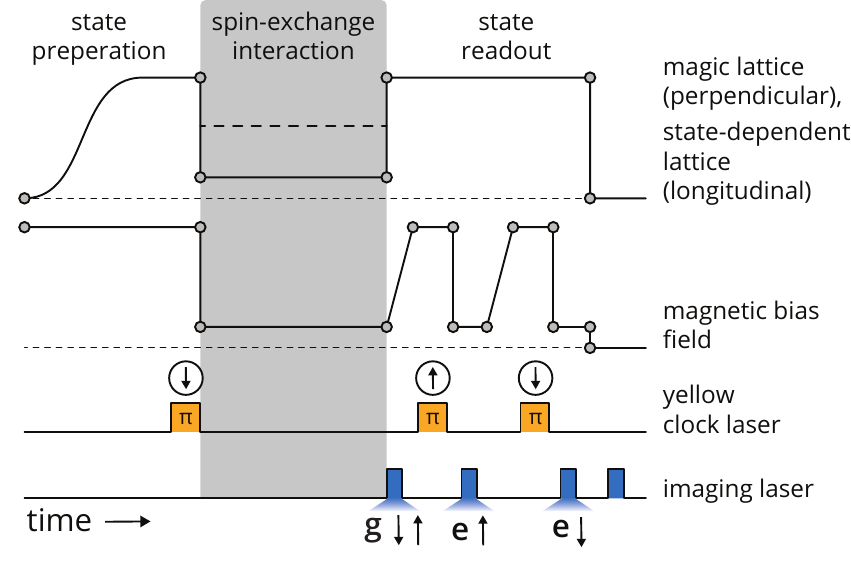}
	\caption{Schematic of experiment sequence after mixture of \gupstate{}- and \gdostate{}-atoms is loaded into the state-dependent optical lattice.}
	\label{fig:setup-sequence}
	\end{centering}
\end{figure}

The degenerate, balanced mixture of \gupstate{}- and \gdostate{}-atoms is prepared by optical pumping into the nuclear spin components $m_F = \pm 5/2$ and subsequent evaporative cooling in a crossed optical dipole trap.
In order to prepare singly occupied lattice sites with either \edostate{} or \gupstate{}-atoms, we first load the two-spin \gstate{}-mixture with relatively low atom numbers of $N=3.3\times 10^3$ per spin state into a 3D lattice ($V_\perp = 45...50 \,E_\mathrm{rec}^\perp$ and $V_z = 8 \,E_\mathrm{rec}^z$). 
We estimate an average filling of around 21 atoms per tube.
The repulsively interacting fermions in the lattice suppress double occupancies. 
We verify this experimentally via clock-line spectroscopy on the initial state, showing no peaks attributable to interacting states. 

A schematic overview of the experimental sequence driving the spin-exchange dynamics is shown in Fig.~\ref{fig:setup-sequence}.
The transfer of \gdostate{} to \edostate{} is done by a $\pi$-pulse on the clock transition. 
A magnetic field of $20\,$G provides enough Zeeman splitting to address the two spin states separately. 
It also suppresses undesired spin-exchange dynamics during the state preparation.
Also, the mobility of the \gstate{}-atoms is reduced by the deep SDL, with a hopping rate of $t_g/h\sim 80\,$Hz.
The $\pi$-pulse is performed at high intensities (at Rabi frequencies of $\Omega = 2\pi\cdot 2.0\,$ kHz) to compensate for spatially varying light shifts of the clock transition due to the harmonic confinement of the SDL beam. 

To initiate the spin-exchange dynamics, the magnetic field is rapidly lowered to a small bias value of $B=1$\,G immediately after the excitation pulse.
The field is high enough to preserve the nuclear spin quantization axis while at the same time inducing only small Zeeman energies ($h\cdot 275$\,Hz) compared to the on-site interaction energies (e.g., $U^- = h\cdot 3.7$\,kHz for $V_\perp = 30 \,E_\mathrm{rec}^\perp$ and $V_z = 6 \,E_\mathrm{rec}^z$). 
Simultaneously, the magic-wavelength lattice and the SDL are ramped, within 1\,ms, to the values $V_{\perp}$ and $V_{z}$, defining the confinement during a variable spin-exchange hold time $\tau$. 

Before separately imaging both excited-state spin components, we image both ground-state spin components by means of a high-intensity imaging pulse. 
This allows us to count the \gstate{}-atoms and also sets a well-defined end point for the spin-exchange dynamics.
We also ramp the lattice depth up again (1\,ms ramp time) in order to suppress further tunneling dynamics during the imaging of \estate{}-atoms. 
To image the \eupstate{}-atoms, we first ramp the magnetic field back up to 20\,G (10\,ms ramp time), spin-selectively deexcite to the ground-state with a $\pi$-pulse and quench the field back to 1\,G. By imaging the deexcited \eupstate{}-atoms they are also removed from the lattice.
This procedure is repeated analogously for the \edostate{}-atoms.
Overall, we are able to count the \gstate{}, \eupstate{} and \edostate{} atom numbers in a single experimental realization.

\section{Lifetime in the SDL}

The enhanced spin-exchange rate on the resonance branches observed in the main text is accompanied by amplified atom losses from the trap. 
A linear fit to the short-time dynamics, up to one tunneling time $h/t_g$ in Fig.~2, yields a maximum initial loss rate of 52\,Hz for the \estate{}-atoms and 6\,Hz for the \gstate-{atoms}.
This indicates that the loss is not entirely caused by an \estate{}-\gstate{}-symmetric process.
	
In an independent measurement, the pure \estate{}-atom losses are quantified.
Therefore, a balanced \eupstate{}-\edostate{}-mixture is prepared at densities comparable to the spin-exchange measurements in the main text. 
The \estate{}-atoms are held in the same lattice configuration as the $eg$-mixture in Fig.~2(a) (black curve) of the main text ($V_z=5.7\,E_\mathrm{rec}^z$, $V_\perp=30.5\,E_\mathrm{rec}^\perp$).
After a time of $10\,h/t_g$, only 6\% of the \estate{}-atoms are lost, 
compared to 35\% in the $eg$-mixture in Fig.~2(a).

On longer timescales, the \estate{}-lifetime in our experiment is limited by vacuum losses and repumping to \gstate{} by the magic-wavelength as well as state-dependent lattice light.
By holding a spin-polarized sample in a 3D isotropic magic-wavelength lattice (30\,$E_\mathrm{rec}$ depth), we measure a loss rate of 28\,mHz induced by vacuum background collisions and an optical repumping rate of 175\,mHz to the ground state, induced by the magic-wavelength light. 

\begin{figure}[htb]
	\begin{centering}
	\includegraphics[width=0.55\columnwidth]{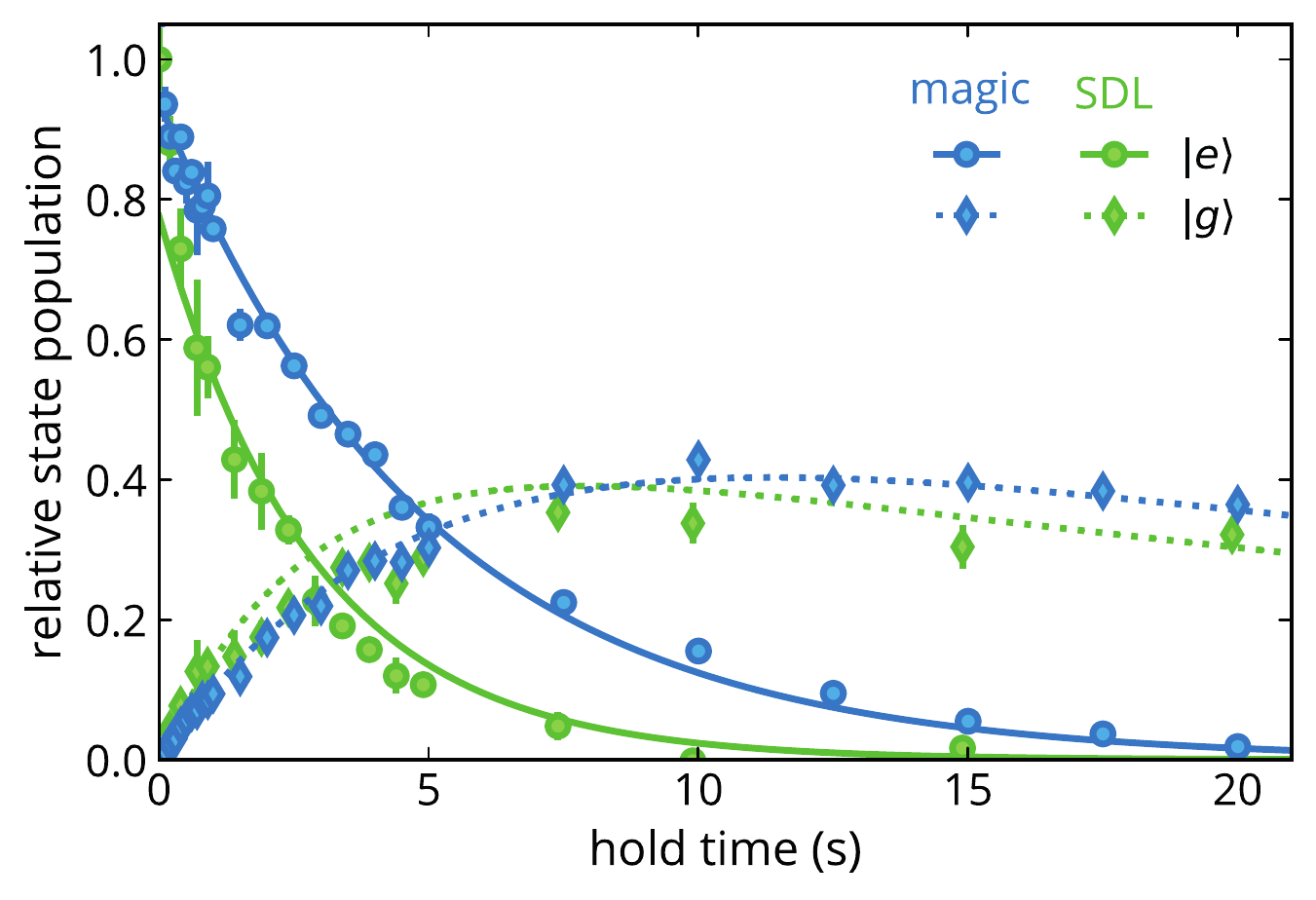}
	\caption{Losses due to repumping by lattice light and background collisions in a 3D lattice: Atom numbers in \estate{} (circles) and \gstate{} (diamonds) for varying hold time in the 30\,$E_\mathrm{rec}$ isotropic magic-wavelength lattice (blue) and in the $5.8\,E_\mathrm{rec}^z$ SDL with 45\,$E_\mathrm{rec}$ perpendicular magic-wavelength confinement (green). Error bars indicate the standard error of the mean. Solid and dashed lines represent fits of the one-body rate equation for the coupled \estate{}-\gstate{} decay dynamics.}
	\label{fig:e-losses}
	\end{centering}
\end{figure}

The longitudinal SDL at a lattice depth of $5.8\,E_\mathrm{rec}^z$ adds an additional repumping rate of 147\,mHz for the \estate{}-atoms. 
This rate was obtained by fitting a rate equation that includes all single-particle loss processes to the \estate{} and \gstate{} atom counts (see Fig.~\ref{fig:e-losses}).
By using 45\,$E_\mathrm{rec}$ deep perpendicular magic-wavelength lattices in addition to the  SDL, the total magic-wavelength light power was maintained compared to the previous measurement. 
The overall lifetime of a spin-polarized \estate{}-sample is 2.8\,s for the lattice configuration specified above, exceeding the experimentally relevant time scales in the main text.

\section{Two particles in mixed anisotropic confinement}
\label{sec:onsite-model}

In order to estimate the on-site energies in the state-dependent lattice, we treat the lattice sites as independent harmonic oscillators.
We consider one \estate{}-atom and one \gstate{}-atom trapped in a mixed confinement, interacting via the regularized contact potential $V_\mathrm{int} = 4\pi a_{eg}^\pm \delta(\vec r) \frac{\partial }{\partial r}{r}$.
While the radial trapping frequency $\omega_{\perp}$ (along the magic) is the same for the two particles, the longitudinal frequencies $\omega_{z,e}$  and $\omega_{z,g}$ (along the SDL) depend on the electronic orbital. 
The total Hamiltonian in cylindrical coordinates ($\rho_{e/g}, z_{e/g}$) is
\begin{equation}
\hat{H}=-\frac{\hbar^{2}}{2m}\left[\mathbf{\nabla}_{g}^{2}+\mathbf{\nabla}_{e}^{2}\right]+\frac{m}{2}\left[\omega_{\perp}^{2}\left(\rho_{g}^{2}+\rho_{e}^{2}\right)+\omega_{z,g}^{2}z_{g}^{2}+\omega_{z,e}^{2}z_{e}^{2}\right]+V_\mathrm{int}(\mathbf{r}_{g}-\mathbf{r}_{e}).
\end{equation}
Let us now use center-of-mass (c.m.) and relative (rel.) coordinates $\mathbf{R}=(\mathbf{r}_{g}+\mathbf{r}_{e})/2$
and $\mathbf{r}=\mathbf{r}_g-\mathbf{r}_{e}$. 
One obtains $\mathbf{\nabla}_{g}^{2}+\mathbf{\nabla}_{e}^{2}=\mathbf{\nabla_{R}}^{2}/2+2\mathbf{\nabla_{r}}^{2}$,
$\rho_{g}^{2}+\rho_{e}^{2}=2\rho_\mathrm{c.m.}^{2}+\rho^{2}/2$, $z_{g}=Z+z/2$
and $z_{e}=Z-z/2$. 
The Hamiltonian becomes
\begin{eqnarray*}
\hat{H} & = & \hat{H}_\mathrm{c.m.}+\hat{H}_\mathrm{rel.}+\hat{H}_\mathrm{mix},\\
\hat{H}_\mathrm{c.m.} & = & -\frac{\hbar^{2}}{2(2m)}\mathbf{\nabla_{R}}^{2}+\frac{2m}{2}\left[\omega_{\perp}^{2}\rho_\mathrm{c.m.}^{2}+\bar{\omega}_{z}^{2}Z^{2}\right],\\
\hat{H}_\mathrm{rel.} & = & -\frac{\hbar^{2}}{2(m/2)}\mathbf{\nabla_{r}}^{2}+\frac{m/2}{2}\left[\omega_{\perp}^{2}\rho^{2}+\bar{\omega}_{z}^{2}z^{2}\right]+V_\mathrm{int}(\mathbf{r}),\\
\hat{H}_\mathrm{mix} & = & \frac{m}{2}Zz\Delta\omega^{2},
\end{eqnarray*}
with an effective longitudinal trapping frequency $\bar{\omega}_{z}^{2}=(\omega_{z,g}^{2}+\omega_{z,e}^{2})/2$. 
The harmonic oscillators in the rel. and c.m. coordinate are coupled by a mixing term $\hat{H}_\mathrm{mix}$ proportional to $\Delta\omega^{2}=\omega_{z,e}^{2}-\omega_{z,g}^{2}$.
In the main text, we use the energy unit $\hbar\bar{\omega}_{z}$, relative coordinate length unit 
$\ell=\sqrt{\hbar/\mu\bar{\omega}_{z}}$ (with $\mu=m/2$ the
reduced mass) and c.m. coordinate length unit 
$L=\sqrt{\hbar/M\bar{\omega}_{z}}$ (with $M=2m$ the
combined mass for the c.m.):
\begin{eqnarray*}
\hat{H}_\mathrm{c.m.} & = & -\mathbf{\nabla_{R}}^{2}/2+\left[\eta^{2}\rho_\mathrm{c.m.}^{2}+Z^{2}\right]/2\\
\hat{H}_\mathrm{rel.} & = & -\mathbf{\nabla_{r}}^{2}/2+\left[\eta^{2}\rho^{2}+z^{2}\right]/2+V_\mathrm{int}(\mathbf{r})\\
\hat{H}_\mathrm{mix} & = & \frac{1}{2}\frac{\Delta\omega^{2}}{\bar{\omega}_{z}^2}Zz,
\end{eqnarray*}
where $\eta=\omega_{\perp}/\bar{\omega}_{z}$ is the anisotropy parameter of the on-site confinement.

For the prediction of the spin-exchange resonance locations in Fig.~3 and Fig.~4 of the main text, we neglect all coupling terms between c.m. and rel. coordinates.
Then, the c.m. problem is simply a non-interacting 3D harmonic oscillator with eigenenergies $E_\mathrm{c.m.}=(\eta+1/2) + n_z + n_\perp \, \eta$.
Here, $n_z$ and $n_\perp$ are the number of longitudinal and perpendicular band excitations. 
Concerning the relative motion, the problem can be solved exactly, as in \cite{idziaszek2005supp}. 
States of the 3D harmonic oscillator with angular momentum $l\neq0$
do not see the contact interaction potential. 
For states with $l=0$, the energy of the relative motion part is $E_\mathrm{rel.}=\eta+1/2+\epsilon$, with
\[
\mathcal{F}(-\epsilon/2;\,\eta)=1/a,\quad\mathcal{F}(x;\,\eta)=-\frac{1}{\sqrt{\pi}}\int_{0}^{\infty}dt\left[\frac{\eta e^{-xt}}{\left(1-e^{-t\eta}\right)\sqrt{1-e^{-t}}}-\frac{1}{t^{3/2}}\right].
\]
The above integral only converges for $x>0$. Yet, the following recursive formula can be used for negative values:
\[
\mathcal{F}(x+\eta;\,\eta)-\mathcal{F}(x;\,\eta)=\frac{1}{\sqrt{\pi}}\int_{0}^{\infty}dt\frac{\eta e^{-xt}}{\sqrt{1-e^{-t}}}=\frac{\eta\Gamma(x)}{\Gamma(x+1/2)}.
\]

To fully treat the on-site interactions, one would need to take into account corrections from the anharmonic part of the lattice potential and the coupling term from the mixed confinement.

\section{Model for spin-exchange tunability}
As a model system for the two-particle spin-exchange dynamics, we consider two particles on two lattice sites (sites $L,R$ in a double well) of the state-dependent lattice. 
One particle is in the excited state $\ket{e}$, the other one is in the ground state $\ket{g}$. 
They can have different spins $\ket{\uparrow}$ or $\ket{\downarrow}$. 
The $\ket{e}$-particle is localized on one of the two sites (site $L$). 
The $\ket{g}$-particle can hop between the two sites with tunnel coupling $t_g$.
Initially, the two particles are prepared on different lattice sites, with a spatial wave function $\ket{C}$.
A spin-exchange process can then only be mediated by tunneling of \gstate{}-atoms.
In contrast, the on-site spin-exchange coupling for doubly occupied sites is given as $V_\mathrm{ex} = (U^+ - U^-)/2>0$ \cite{Scazza2014Supp, Cappellini2014Supp}.

Considering only the typical lowest-band Hubbard parameters, the interorbital interaction strengths $U^-$ and $U^+$ for the spin-triplet and spin-singlet interaction channels are both positive and large compared to the hopping rate $t_g$.
For our experiment, this is true everywhere in the tight-binding regime for both interaction channels. 
In the sampled range of $V_z = (3 ... 8)E_\mathrm{rec}^z$, we obtain $4t_g/U^- = (0.52 ... 0.10)$ at $V_\perp=30 E_\mathrm{rec}^\perp$.
Spin exchange then happens only via a virtual process and the effective coupling scales as $J_\mathrm{ex}^-\propto t_g^2/U$.
Mainly, the lower-energy spin triplet contributes to the exchange process, leading to a ferromagnetic coupling
\begin{equation}
J_\mathrm{ex} = - \frac{t_g^2}{U^-} + \frac{t_g^2}{U^+} \approx - \frac{t_g^2}{U^-} < 0 \, 
\end{equation}
in the low-energy limit. 

As explained in the main text and Sec.~\ref{sec:onsite-model}, we expect a coupling of c.m. and rel. motion that can lead to two particles on separate sites coupling into a c.m.-exited on-site bound state.
The on-site c.m. excitation is then labeled by $\ket{B}$. The tunnel coupling from neighboring sites into that state is defined as $t_b$.

If two particles tunnel onto the same site ($L$) and go into an on-site wave function $\ket{A}$ without c.m. excitation, the possible interaction channels are $U^-$ and $U^+$, depending on whether they are in the spin triplet $\ket{-} = \frac{1}{2}(\ket {ge} - \ket{eg}) (\ket{\uparrow\downarrow} + \ket{\downarrow\uparrow}) \ket{A}$ or spin singlet 
	$\ket{+} = \frac{1}{2}(\ket {ge} + \ket{eg}) (\ket{\uparrow\downarrow} - \ket{\downarrow\uparrow})\ket{A}$.

If two particles tunnel onto the same site while acquiring a band excitation into $\ket{B}$, they will dominantly interact via a c.m.-excited, spin singlet/triplet bound state close to the entrance energy, as described in he main text.
The total on-site energy, considering also the c.m. excitation, is then $U^\pm_{b}$ for the states $\ket{\pm_b} = \frac{1}{2}(\ket {ge} \pm \ket{eg}) (\ket{\uparrow\downarrow} \mp \ket{\downarrow\uparrow})\ket{B}$.
The interaction energies $U^\pm$ and $U^\pm_{b}$ are from the on-site interaction model in Sec.~\ref{sec:onsite-model}.

The corresponding Hamiltonian is:
\begin{eqnarray*}
H = &-t_g \sum_\sigma \left(\ket{g,\sigma}\ket{A} \bra{C} \bra{g,\sigma}+ h.c.\right) \\
    &-t_b \sum_\sigma \left(\ket{g,\sigma}\ket{B} \bra{C} \bra{g,\sigma}+ h.c.\right) \\
    &+ U^- \ket{-}\bra{-} + U^+ \ket{+}\bra{+} \\
    &+ U^+_b \ket{+_b}\bra{+_b}+ U^-_b \ket{-_b}\bra{-_b}.
\end{eqnarray*}
In order to study the effect of bound-state coupling onto the spin-exchange dynamics, study the time evolution of the Hamiltonian. 
For computation we use a product state basis.
With internal states $\ket{\mathrm{orbital, spin}}$ and the spatial wave function $\ket{\mathrm{A/B/C}}$, the relevant two-particle basis states are $\ket{e,\uparrow} \ket{g,\downarrow} \ket{A}$, $\ket{e,\downarrow} \ket{g,\uparrow} \ket{A}$, $\ket{e,\uparrow} \ket{g,\downarrow} \ket{B}$, $\ket{e,\downarrow} \ket{g,\uparrow} \ket{B}$, $\ket{e,\uparrow} \ket{g,\downarrow} \ket{C}$ and $\ket{e,\downarrow} \ket{g,\uparrow} \ket{C}$ (we omit the trivial fermionization).
In this basis, the two-site model has the following form:
\begin{equation}
	\hat H  = \left[ \begin{array}{cccccc}
	\frac{U^-+U^+}{2} & \frac{U^--U^+}{2} & 0 & 0 & -t_g & 0 \\
	\frac{U^--U^+}{2} & \frac{U^-+U^+}{2} & 0 & 0 & 0 & -t_g \\
	0 & 0 & \frac{U^-_b+U^+_b}{2}& \frac{U^-_b-U^+_b}{2}& -t_b & 0 \\
	0 & 0 & \frac{U^-_b-U^+_b}{2}& \frac{U^-_b+U^+_b}{2}& 0 & -t_b \\
	-t_g & 0 & -t_b & 0 & 0 & 0 \\
	0 & -t_g & 0 & -t_b & 0 & 0 \\
	\end{array} \right]\, .
\end{equation}

\begin{figure}[tb]
	\begin{centering}
	\includegraphics[width=0.8\columnwidth]{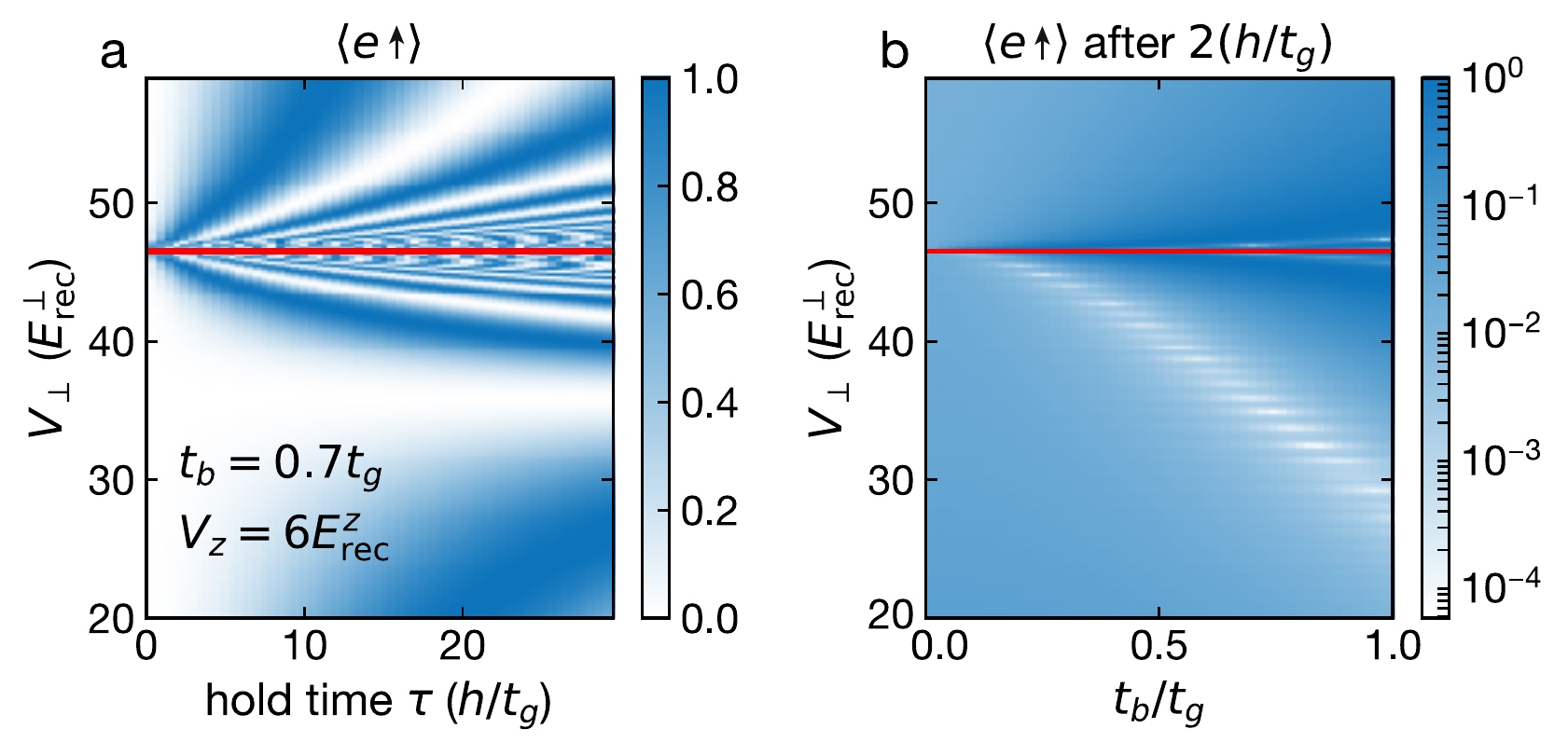}
    \caption{Spin-exchange dynamics from state-dependent double-well model (sites $L,R$):
    (a) Expectation value of repopulated $e\uparrow$-fraction after initial state $\ket{e\downarrow}_L\ket{g\uparrow}_R$ is held for $\tau$ at coupling strength $t_b = 0.7t_g$ and different perpendicular confinements. Around the resonance position (solid red line), where $\ket{+_b}$ comes into resonance with the entrance energy $E_0$ of two particles on different sites, fast coherent oscillations with frequency $\sim t_b$ occur. 
    (b) Dependence of the dynamics on the coupling strength $t_b$ of the initial state into the c.m.-excited singlet bound state $\ket{+_b}$. 
    Confinement dependent expectation value of repopulated \eupstate-fraction after a short hold time $\tau = 2\,h/t_g$.
	}
   	\label{fig:double-well-dynamics}
   	\end{centering}
\end{figure}

Fig.~\ref{fig:double-well-dynamics}(a) depicts the coherent spin-exchange dynamics happening in the double-well model for different confinements at a finite coupling strength into the c.m.-excited bound state.
The data is centered around the energetic resonance of the second c.m. excitation  of the bound spin-singlet state $\ket{+_b}$ with the entrance energy $E_0$.
The \eupstate{}-state is repopulated through spin-exchange with a rate that shows a resonance structure in the confinement similar to the one observed experimentally (see main text). 
In Fig.~\ref{fig:double-well-dynamics}(b), the influence of the coupling strength on the width of the resonance is illustrated.
On the low-confinement side of the resonance, the contributions of the intermediate states $\ket{+_b}$ and $\ket{-}$ to the effective spin-exchange coupling cancel and lead to a zero in the exchange rate.

This model is based on the decoupled harmonic-oscillator provided in Sec.~\ref{sec:onsite-model} and therefore does not incorporate on-site coupling between c.m. and rel. coordinate, potentially leading to further coupling terms and energetic shifts of the Hubbard parameters.

\end{document}